\documentclass[useAMS,usenatbib]{mn2e} 
\usepackage{graphicx,epsf,ifthen} 
\newcommand{\figtype}{EPS} 
\def\myputfigure#1#2#3#4#5%
{\vskip#5pt\makebox[0pt]{\hskip#2in 
\includegraphics[width=#3\textwidth]{#1}}\vskip#4pt\hfill}

\title[Morphological redshifts]{Morphological Redshift Estimates for  
Galaxy Clusters in a Sunyaev-Zel'dovich Effect Survey} 
 
\date{draft} 
 
\author[J.M Diego et al.] 
   { J.M  Diego$^1$, J. Mohr $^2$, J. Silk$^1$, and G. Bryan$^1$. \\ 
   $^1$Astrophysics Dept. Univerity of Oxford. Denys Wilkinson building. 1, Keble Road, Oxford OX1 3RH,  UK\\  
   $^2$Departments of Astronomy and Physics, University of Illinois, Urbana, IL 61801.} 
 
\pagerange{\pageref{firstpage}--\pageref{lastpage}} 
 
\begin{document} 
 
\maketitle 
 
\label{firstpage} 
\begin{abstract} 
 
We develop a new method to estimate the redshift of galaxy clusters  
through resolved images of the Sunyaev-Zel'dovich effect (SZE). Our method is based on morphological  
observables which can be measured by actual and future SZE experiments.    
We test the method with a set of high resolution hydrodynamical simulations  
of galaxy clusters at different redshifts.  
Our method combines the observables in a principal component analysis.  
After {\it calibrating} the method with an independent redshift estimation for some of the  
clusters, we show-- using a Bayesian approach-- how the method can give an estimate of the  
redshift of the galaxy clusters. 
Although the error bars given by the morphological redshift estimation are large,  
it should be useful  
for future SZE surveys where thousands of clusters are expected to be detected;  
a first preselection of the high redshift candidates could be done using our proposed morphological  
redshift estimator. 
Although not considered in this work, our method should also be useful to give an estimate of the  
redshift of clusters in X-ray and optical surveys. 
 
\end{abstract} 
 
 
\section{Introduction} 
\label{section_introduction} 
 
The advent of new experiments dedicated to the observation of the Sunyaev-Zel'dovich  
effect (Sunyaev \& Zel'dovich, 1972) (SZE hereafter),  
demands the development of new techniques to best analyze  
these new and exciting data. 
With the SZE it is possible to probe the hot plasma in  galaxy clusters, which  
shifts the spectrum of the cosmic background radiation.  
This shift is redshift independent, and it is proportional to the temperature of  
the plasma and its electron density ($n_e$). This characteristic ($z$-independent, and $\propto n_e$)  
makes the SZE an ideal way to explore the high redshift population of galaxy clusters. 
 
However, the fact that the SZE distortion is independent of the redshift  
of the cluster makes the determination of the redshift  
of the cluster a challenging task.  
Redshift information is crucial if one attempts to use cluster surveys to study the  
evolution of our universe.   
The evolution of the cluster number counts ($dN/dz$) is a very sensitive indicator  
of the cosmological model \citep{eke98,mathiesen98,henry00}.  
The local abundance of clusters shows a degeneracy  
in $\Omega -- \sigma_8$ \citep{eke96,bahcall97} but this degeneracy can be broken with an accurate  
estimation of $dN/dz$ up to moderate or high redshifts ($z \approx 0.5 - 1$)  
\citep[see e.g][]{Bahcall98, borgani01}.  
The cluster redshift distributions in suitably large SZE cluster surveys  
can potentially provide precise constraints on the amount and nature of the dark energy in the  
universe \citep{haiman01,holder01b,weller01,majumdar02}. 
Redshifts are also necessary to study the evolution of the cluster structure and dynamics. 
 
One normally determines the redshift using photometric and spectroscopic observations  
of the galaxies in the cluster.  
Spectroscopic observations of galaxies in relatively nearby clusters are straightforward, but for distant  
clusters it is challenging even with the largest telescopes.   For large solid angle surveys, photometric  
redshifts will be of critical importance, allowing redshift determination for far less time invested at  
the telescope.  However, photometric redshifts  
are also time consuming and for clusters above redshift $\approx$1, photometric redshifts require large  
telescopes (see for instance Diego et al. 2002 where the authors show the selection function for  
a galaxy cluster survey with a 10-m telescope and photometric redshift estimations). 
Future SZE experiments will detect hundreds and perhaps thousands of galaxy clusters.  
The {\it Planck} Surveyor SZE survey is expected to detect more than 10$^4$ clusters with  
redshifts extending to $\sim$2 \citep[depending on the cosmological model,][]{diego02}.  A planned,  
arcminute resolution SZE survey from the South Pole will detect similar numbers of clusters with a  
much larger fraction at high redshift. 
 
Measuring redshifts for large solid angle, high redshift cluster surveys is a daunting task.    
An optimal solution may be to combine small and medium--sized telescopes to determine the redshifts of  
the low and intermediate $z$ clusters, reserving the redshift measurements of the most distant clusters  
for the largest available telescopes.  Clearly this strategy requires crude {\it a priori} knowledge of  
the cluster redshifts. The motivation of the work described here is to examine whether it is possible to  
make this preselection of the low, intermediate and high redshift clusters using SZE data alone. 
 
Our method is based only on observed SZE properties of galaxy clusters.   
These include  the observed shape and size of the cluster, which do have some  
dependence on the redshift.  For instance, the apparent size of a particular cluster will  
decrease when increasing its redshift.  So, an apparent size will, in principle,  
constrain the cluster redshift.  However, the apparent size of the cluster also depends on its total mass.  
Two clusters with different redshifts and masses can have the same apparent  
size, provided the more distant cluster has a larger mass that exactly compensates for the decrease in the  
apparent size due to the increased redshift. There is, therefore, a degeneracy between  
the cluster redshift and mass.  
 
The question is whether we can break  
this degeneracy by using additional information. A resolved SZE image of a cluster  
provides information not only about the cluster size, but also about the shape of the cluster gas  
distribution. The total observed flux of the cluster, for instance, depends on the total cluster mass,  
the redshift and the temperature. The central SZE decrement depends  
on the core radius and the electron central density, but it is, in principle, independent of the redshift.  
Our method incorporates these and other observables to break the mass--redshift degeneracy.  This method  
requires resolved SZE images.  
Therefore, it should be useful for arcmin and sub-arcmin resolution experiments but not for experiments  
like {\it Planck},  where the best resolution will be 5~arcmin. 
 
In this work we will not consider the effects of the relativistic corrections  
and the kinematic effect, because they are small compared with the non-relativistic  
thermal SZE. Their effect will be discussed in a later paper.  In $\S$\ref{sec:clusters} we outline the  
connections between cluster morphology and redshift from a structure formation viewpoint.   
$\S$\ref{sec:method} discusses some of the weaknesses of this theoretical perspective and then  
provides a detailed description of a method that overcomes these weaknesses.  A demonstration of the  
degree to which the method works is contained in $\S$\ref{sec:application}, and a discussion of  
conclusions follows in $\S$\ref{sec:conclusions}. 
 
\section{Connections Between Cluster Morphologies and Redshifts} 
\label{sec:clusters} 
In this section we will use theoretical arguments to justify the use of morphological redshifts.  
We will  develop a simple analytic cluster SZE model and then use it to compute cluster observables.   
For the particular case of the isothermal $\beta$-model in a cosmological model  
with $\Omega _m = 1$, we will demonstrate in subsection 2.2 how combined measurements of cluster 
size and flux lead directly to a redshift estimate.  Although this cosmological model is not consistent with  
current data, it's simplicity make it useful for illustrating the method, and the results are fully  
generalizable. 
 
\subsection{A Model for Cluster SZE Signatures} 
The distortion in the CMB intensity due to thermal SZE is 
\begin{equation} 
\Delta I = I_o*f(x)*y_c 
\label{eq_DeltaI} 
\end{equation} 
where $I_o \approx 2.7 \times 10^{11} \frac{mJy}{sr}$, $x$ is the adimensional frequency 
($x = h\nu/k_b T \approx \nu (GHz)/56.8$), f(x) is the frequency dependence of the  
SZE $f(x) = [x coth(x/2) - 4]\times [x^4 e^x /(e^x - 1)^2]$ and  $y_c$ is the cluster 
Compton parameter: 
\begin{equation} 
y_c = \frac{k_b \sigma _T}{m_e c^2}\int T n(l) dl. 
\label{eq_yc} 
\end{equation} 
This distortion is independent of cluster redshift because  
$x$ does not depend on the redshift (because both $\nu$ and $T$ depend on the  
redshift in the same way).  Therefore, the SZE spectral distortion  
provides no redshift information about the cluster. 
 
However, if we can produce a resolved SZE image of a cluster, we gain much more information and can  
potentially solve for the redshift. To illustrate  
this point, let us assume the simple case of a cluster at redshift $z$ with an  
electron density profile described by a $\beta$-model \citep{cavaliere78} with $\beta = 2/3$, which is  
the value found to best match clusters \citep{jones84,mohr99}.  In this case the electron density  
profile is just $n( r ) = n_o/\left(1 + (r/r_c)^2\right)$,  
where $n_o$ is the electron central density and $r_c$ is the core radius, which we take to be some  
constant fraction of the virial radius $r_c=r_v/p$.  $p$ is a parameter with value ranging between 10  
and 20.  More realistic modelings that included, for example,  a mass dependent $p$, could be  
considered, but for illustrative purposes our simple model will suffice. 
 
The redshift evolution of these parameter can be modeled as  
\begin{equation} 
n_o = N_o \frac{\Delta _c(z)}{\Delta _c(0)}E^2(z) 
\end{equation} 
where $\Delta _c(z)$ is the critical collapse overdensity with respect to the critical  
density at redshift z ($\Delta _c(z) = 18\pi^2 + 82x -39 x^2$ with $x = \Omega(z) - 1$ and 
$\Omega(z) = \Omega_m(1 + z)^3/E(z)^2$)
and $H(z) = H_oE(z)$ \citep[see e.g][]{bryan98,mohr00a}.  
$N_o$ is adjusted to fix the total gas mass  
$M_{gas}=\mu _e m_p \int_o^{r_v} n(r) 4\pi r^2 dr \approx f_b M$ 
where $M$ is the virial mass and $f_b$ is the baryon fraction.   
For a fully ionized, purely hydrogen gas $\mu _e = 1$. 
Similarly, for the virial radius we have  
\begin{equation} 
r_v = R_o M^{1/3} \left( \frac{\Delta _c(z)}{\Delta _c(0)}E(z)^2 \right) ^{-1/3}. 
\label{eq_rv} 
\end{equation} 
Using the virial theorem ($2K + V = 0$) and the spherical collapse model, 
we can obtain an expression for the virial temperature of the cluster.  
\begin{equation} 
T = T_o M_{15}^{2/3}\left( \frac{\Delta _c(z)}{\Delta _c(0)} E(z)^2 \right) ^{1/3}. 
\label{eq_T} 
\end{equation} 
The normalization, $T_o$ can be obtained from models or from a fit of this 
relation to the data. We will adopt the second approach and use the values derived 
in \citep{diego01}.   
 
Within this model, the Compton parameter in the direction $\theta$ is  
\begin{equation} 
y_c(\theta) =  \frac{k_b \sigma _T}{m_e c^2} r_c n_o T \Phi(\theta) = y_o \Phi(\theta) 
\label{eq_yc_theta} 
\end{equation} 
where $\theta$ is the angle between the line of sight and the center of the cluster.  We have assumed  
here that $T$ is constant, and we have ignored any contributions to the $y_c$ from outside the cluster  
virial region.  
The function $\Phi(\theta)$ is just the integral of the density profile along the line of  
sight. 
\begin{equation} 
\Phi(\theta) = \frac{2}{\sqrt{1 + (\theta/\theta _c)^2}} tan ^{-1}  
\sqrt{\frac{p^2 - (\theta/\theta _c)^2}{1 + (\theta/\theta _c)^2}} 
\label{eq_Phi} 
\end{equation} 
$\theta _c$ is the apparent core radius $r_c/d_A$.  
The cluster surface brightness profile is then 
\begin{equation} 
B(\theta) = B_o r_c n_o T \Phi(\theta)  
\label{eq_2DB} 
\end{equation} 
where (see Eqn ~\ref{eq_DeltaI}) 
$B_o = I_o f(x) \left(k_b \sigma _T/m_e c^2\right)$.  
Using this model we will now show how cluster  
morphologies contain information about the cluster redshift. 
 
\begin{figure} 
   \ifthenelse{\equal{\figtype}{EPS}}{ 
   \begin{center} 
   \epsfxsize=8.cm  
   \begin{minipage}{\epsfxsize}\epsffile{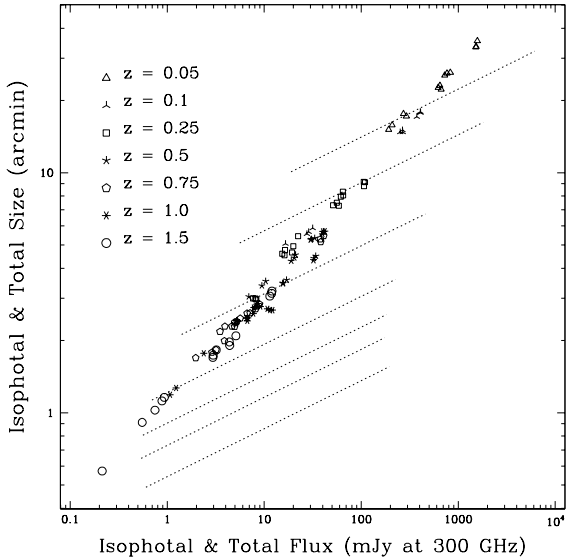}\end{minipage} 
   \end{center}} 
   {\myputfigure{f1.pdf}{3.2}{0.55}{-180}{-30}} 
   \caption{\label{fig_S_D_ideal} 
            Ideal vs real case. The dotted lines represent the expected cluster total size-flux relation 
            in the ideal case where the cluster is observed with infinite resolution and sensitivity.
            Each line corresponds to a different redshift (listed on the left) for a range of masses 
            $3\times 10^{13}-1\times 10^{15} h^{-1} M_{\odot}$. 
            Top line is for $z=0.05$ while bottom line is for $z=1.5$. 
            The symbols show the case in {\it real life} (simulations) where one observes isophotal 
            quantities and the scaling relations differ from the case of the toy model. 
            Note that in this case, the characteristic mass as a function of redshift for 
            the simulated clusters is decreasing with redshift. This, and the fact that 
            we are measuring isophotal sizes instead of virial sizes, 
            explains the shift to the left with redshift.
           } 
\end{figure} 
 
\subsection{Redshifts from Total Flux and Apparent Size in the $\Omega = 1$ Case.} 
We now apply the model developed above to demonstrate that in principle two simple observables provide  
enough information to estimate cluster redshifts.  We specifically use the evolution model appropriate  
for  $\Omega_M=1$ only for simplicity. In this simple case, the general expressions in the previous 
section reduce their complexity since $\Delta _c(z) = \Delta _c(0) = 18\pi^2$ and $E(z)^2 = (1+z)^3$.
 
The total cluster flux is just  
the integral of the SZE distortion or surface brightness over the entire solid angle of the cluster. 
$S = \int B(\theta) d\Omega$.  Taking a step back and noting that the surface brightness is a line  
integral, it is clear that the total SZE flux is simply an integral over the cluster volume.   
Writing the volume element $dV(z)=d\Omega d_A(z)^2 dl$, where $d_A$ is the angular diameter  
distance, we then show that 
$S=\int d\Omega \int dl\, T n  = d_A^{-2} \int dV\, T n \propto \bar{T} M_{gas} d_A^{-2}.$ 
The total cluster flux is an interesting quantity, depending on the density weighted temperature 
$\bar{T}$, the total gas mass (but not its detailed distribution) and the cluster distance.  
In the isothermal case the total flux at the frequency $x$ is   
$S(x) = S_o\left(f_b T M_{15}\right)/d_A^2,$ 
where $S_o\approx 3.781\times f(x)$ mJy. We take the baryon fraction to be consistent with  
SZE observations \citep[$f_b\approx0.08h^{-1}$][]{grego01}. The mass $M_{15}$ is expressed 
in units of $10^{15} h^{-1} M_{\odot}$. In these units, the $h^{-2}$ dependence of $d_A^2$ 
is cancelled with the $h$-dependence of $f_b$ and $M_{15}$ making the flux $h$-independent.
If we substitute mass for temperature (equation \ref{eq_T}), we end up  
with an expression which only depends on the mass and the redshift. 
\begin{equation} 
S = \frac{S_o T_o f_b M_{15}^{5/3}(1 + z)}{d_A^2}. 
\label{eq_Flux_B}\end{equation} 
The total flux of the cluster depends on its redshift through the angular  
diameter distance and inherent evolution of cluster structure. 
 
\begin{figure} 
   \ifthenelse{\equal{\figtype}{EPS}}{ 
   \begin{center} 
   \epsfxsize=8.cm  
   \begin{minipage}{\epsfxsize}\epsffile{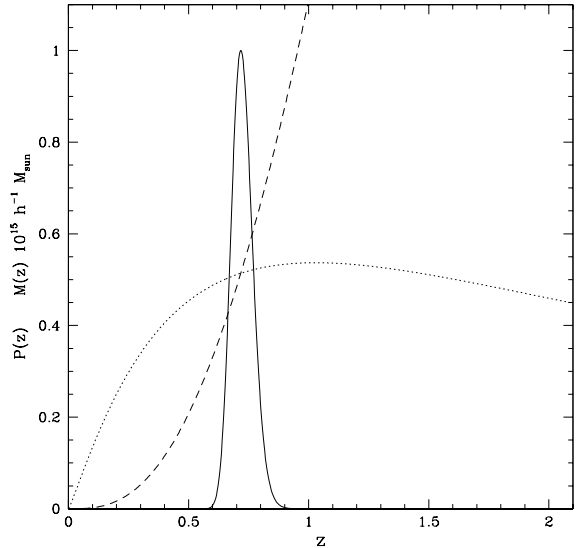}\end{minipage} 
   \end{center}} 
    {\myputfigure{f2.pdf}{3.25}{0.50}{-85}{-30}} 
  \caption{\label{fig_z_recov_perfect} 
            The ideal case. If the clusters follow exactly the scaling relations of  
            equations (\ref{eq_Flux_B}) and (\ref{eq_Diam}), then the recovered redshift  
            could be as good as the one shown in this figure.  
            The solid line shows the Gaussian pdf of the recovered redshift for an  
            experiment with 10 arcsec FWHM. In the computation of the pdf   
            we have assumed that the flux is measured with no error 
            and the diameter has an uncertainty given by the FWHM of the experiment.
            The dotted line is the mass as a function of redshift for a given  
            flux, and the dashed line is the same for a given diameter.  
            This plot is for a cluster with an apparent radius of 4 arcmin and a  
            total flux of 100 mJy (at 353 GHz).   
           } 
\end{figure} 
 
The apparent size of the cluster $\theta_{cl}$ is another quantity which strongly depends on  
the cluster redshift.   If we consider that the physical size of the cluster is related to its 
virial radius, then its observed apparent size is  
(see Eqn.~ \ref{eq_rv}): 
\begin{equation} 
\theta_{cl} = \frac{2 R_o M^{1/3}}{d_A(1+z)} 
\label{eq_Diam} 
\end{equation} 
where a cluster of virial mass $M = 1\times 10^{15} M_{\odot}$ has virial radius $R_o$ Mpc at redshift  
$z=0$.  
If we compare Eqns~\ref{eq_Flux_B} and \ref{eq_Diam}, we can see that both depend  
only on the redshift and cluster mass  (assuming that the other parameters $S_o, T_o, f_b $ and $R_o$ 
are fixed by local observation).  
Thus, in principle, measurements of total flux $S$ and apparent size $\theta_{cl}$ provide both the  
cluster mass and redshift. 
 
Fig. \ref{fig_S_D_ideal} contains a plot of the correlation between total flux and 
apparent size at different redshifts using the model developed above.  
Each line in this plot represents a single redshift and range of cluster mass  
between $3 \times 10^{13} h^{-1} M_{\odot}$ (left) and  
$1.0 \times 10^{15}  h^{-1} M_{\odot}$ (right).  
The redshifts are, from top to bottom, $z = 0.05, 0.1, 0.25, 0.5, 0.75, 1, 1.5$.  
This plot shows clearly that different redshifts are well separated, allowing one to solve for  
redshift and mass with observations of the total flux and apparent size.  
The symbols show the case of simulated data. In this case, the clusters do not follow the scaling 
relations of the previous model. 
Note that the sizes of the simulated clusters are isophotal sizes
rather than virial sizes, because there is no clear observational
signature of the virial region in the SZE properties of a cluster.
 
Both Eqn.~\ref{eq_Flux_B} \& \ref{eq_Diam} can be solved for mass $M$.  Figure~\ref{fig_z_recov_perfect}  
contains a plot of $M$ versus $z$ for specific values of the total flux $S$ (dotted line) and the  
apparent size $\theta_{cl}$ (dashed line).  These two functions intersect at the real redshift of the  
cluster.  This intersection along with the measurement of cluster size could be used to produce a  
probabilistic statement about the cluster redshift.  This is illustrated by the solid line in  
Fig~\ref{fig_z_recov_perfect}, which shows a Gaussian probability distribution for the cluster  
redshift that reflects the uncertainties in the measured apparent size. These uncertainties are  
modelled as a function of the spatial resolution of the experiment (FWHM) which introduces an 
uncertainty in the observed apparent size and consequently on the derived redshift. 
 
\section{A Method for Estimating Redshifts} 
\label{sec:method} 
 
The arguments outlined in the previous section are idealized. In a real experiment, 
the situation would depart from that described above for several reasons: 
(i) clusters do not follow the scaling relations of Eqns.~\ref{eq_Flux_B}  
and \ref{eq_Diam} perfectly, because of departures from equilibrium and variation in cluster structure  
due to ongoing merging that is quite common \citep[e.g.][]{mohr95}, (ii) a real experiment is affected by  
instrumenatl noise and limited by sensitivity, which limits ones ability to estimate the total flux  
and apparent size.   Estimating cluster redshifts from SZE observations is then a much more  
complicated task in practice (see symbols in Fig. \ref{fig_S_D_ideal}); nevertheless, 
the underlying scaling outlined in the previous section is  expected to be a good description of the 
cluster population in a statistical sense.  Therefore, here we  
describe an empirical method for estimating redshifts using SZE morphology and calibration through direct  
redshift measurement in a subsample of the clusters. 
 
In developing this method we are guided by several critical realizations: 
\begin{itemize} 
\item In the ideal case we have assumed that {\it all} clusters  
  lie perfectly on self--similar scaling relations (Eqns~\ref{eq_Flux_B} and \ref{eq_Diam}). 
 There are observed scaling relations in the galaxy clusters which connect, for example, X--ray  
luminosity to  temperature and  virial mass to temperature.  But even in  
  these two well known cases, the scaling relations have an intrinsic scatter,  
  and there is still an ongoing debate about the exact form of these relations.  Thus, any redshift  
estimator that employs scaling relations must allow for scatter and must not require that the exact  
form of the scaling relation be known. 
 
\item Due to noise sources and the limited instrument sensitivity, it will not be possible  
  to observe the entire extent of a cluster. Therefore, it will be difficult to estimate  
  the total flux and size of the cluster from the observed signal.   
  One alternative is to work directly with the observed quantities 
  like the isophotal flux and isophotal size, where the isophote is chosen to lie well above the noise  
limits of the data.   Thus, our method must work with readily available observational quantities. 
 
\item X--ray observations indicate that cluster gas distributions can be reasonably well approximated  
with $\beta$--models, but important  departures remain \citep[e.g.][]{mohr99}.  Observations also  
indicate that clusters are not isothermal \citep[e.g.][]{markevitch98}, which is no surprise given  
the prevalence of merging (and$/$or temperature gradients). 
Thus, our method will have to allow for the fact that cluster structure varied significantly 
from system to system. 
 
\item In practice, observations may provide significantly more information than is contained in the  
isophotal flux or size alone.  Therefore, we need to develop a method that can handle multiple  
observables-- even redundant observables-- in an optimal and graceful manner. 
\end{itemize} 

\subsection{SZE Observables}\label{section_SZE_Observables}
The idea behind morphological redshift estimation is that by combining many observables  
taken from the 2D SZE cluster profile it is possible to {\it divide} the clusters in  
different groups, each one for a different redshift interval. 
The observables must be such that they take into account the last points of the previous  
section.  The list below is not exhaustive, but it includes all the observables used in the following  
section in our attempt to estimate cluster redshifts. 
 
\begin{figure} 
   \ifthenelse{\equal{\figtype}{EPS}}{ 
\begin{center} 
   \epsfxsize=8.cm  
   \begin{minipage}{\epsfxsize}\epsffile{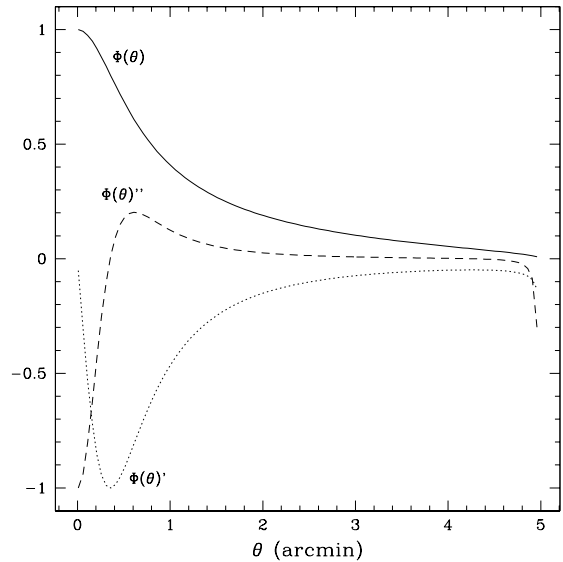}\end{minipage} 
   \end{center}} 
   {\myputfigure{f4.pdf}{3.25}{0.50}{-85}{-30}} 
   \caption{\label{fig_profile} 
           Projected density profile (solid line) as a function of radius. Also shown are the  
           first derivative (dotted line) and the second derivative (dashed line). All curves  
           have been renormalized. The core radius is $\theta _c = 0.5$ arcmin and the ratio  
           between the virial and core radius is $p = 10$. 
           } 
\end{figure}

\begin{itemize} 
\item {\bf Isophotal size}. 
The apparent isophotal size (mean diameter) $\theta_I$ is given by the following expression 
\begin{equation} 
\theta_I = 2 \sqrt{A/\pi} 
\end{equation} 
where $A$ is the total (apparent) area enclosed by the isophote \citep{mohr97a}. In this work, we will use  
an isophote defined to be well above the instrument sensitivity. This sensitivity defines a  
threshold in the 2D images.  The X--ray isophotal size exhibits a tight correlation with the  
emission weighted X--ray temperature both in local samples \citep{mohr97a} and in intermediate  
redshift samples \citep{mohr00a}.  Current SZE observations do not have the required sensitivity  
to examine this property, but many future experiments will have sufficient sensitivity. 
 
\item{\bf Isophotal flux}. 
The isophotal flux is just the total flux within the isophote.   
\begin{equation} 
S_I = \int_A S(\theta) d\Omega 
\end{equation} 
This quantity has never been examined in X-ray observations of ensembles of clusters, but in  
hydrodynamical simulations this quantity appears to be strongly correlated with the isophotal size. 
Nevertheless, we include this quantity, because it may provide additional information, and our  
method handles redundant observables gracefully. 
 
\item {\bf Central amplitude}.  
For the case of the $\beta$-model described above the central amplitude is:  
\begin{equation}          
A_o = B_o r_c n_o T \Phi(0)      
\label{eq_Ao} 
\end{equation}  
The central amplitude only depends on the redshift through the intrinsic cluster  
evolution ($z-$dependence of $r_c$, $n_0$ and $T$). 
The central amplitude is the integrated effect of the projected electron  
population through the cluster center.  If one assumes the scaling relations given in Eqns~\ref{eq_T} and  
\ref{eq_rv}, the central decrement is directly proportional to  
the total cluster mass.   The situation will be more complicated in real clusters, but the example of the  
$\beta$-model is useful to illustrate the utility of the central decrement.  
 
\item {\bf First and second derivatives of the SZE profile}. 
The first and second derivatives of the observed brightness profile for the $\beta$-model  
(see Eqns~\ref{eq_Phi} and \ref{eq_2DB})  
are shown in Figure \ref{fig_profile}, where we also show the projected density  
profile (proportional to $\Phi(\theta)$) for comparison.   
The curves have been renormalized by their respective maximums.  
The core radius in this case was $\theta _c = 0.5$~arcmin, and $p=10$. 
 
The second derivative evaluated at the cluster center ($\theta = 0$) is 
\begin{equation} 
\frac{d^2 B(\theta)}{d\theta^2}(0) = - B_o \frac{r_c n_o T}{\theta _c^2}  
\left( \Phi(0) + \frac{2}{p} \right )   
\label{eq_Deriv2nd_0} 
\end{equation} 
which depends on the redshift through the evolution of $\theta _c$. 
In the regime of interest ($p>>2$), the second derivative in the center is a tracer of the core radius. 
 
The first derivative in the center is null for all the clusters, but it is clear in Fig~\ref{fig_profile} 
that the first derivative reaches a maximum at  $\theta\sim\theta _c$.  The position of the maximum  
coincides with the region where the second derivative vanishes.  The value of the first derivative  
at the core radius is: 
\begin{equation} 
\frac{d B(\theta)}{d\theta}(\theta _c) \propto  \frac{B_o r_c n_o T}{\theta _c} = B_o n_o T d_A 
\end{equation} 
That is, the first derivative in the region where the second derivative vanishes is independent   
of the core radius and is proportional to the angular diameter distance. 
In cases where the core radius is proportional to the mass, the second  
derivative in the center and the first derivative in the region where the second derivative vanishes  
should be useful in breaking the degeneracies between the cluster mass   
and redshift. 
 
\item {\bf Minkowski functionals}. 
Several recent works have suggested that Minkowski functionals applied to galaxy clusters should  
be good tracers of the cluster evolution \citep{beisbart01}. If clusters form by merging  
events, their internal structure should evolve with redshift (although local clusters are known to  
exhibit lots of evidence for merging).  Morphological evolution can in principle be traced with  
Minkowski functionals.  
In this paper we will consider three of them: (i) total perimeter of the isophote,  
(ii) ellipticity of the isophote and (iii) number of subgroups  
above the isophote. In a recent paper employing Minkowski functionals, the authors claim  
that some evolution with redshift in the ellipticity of galaxy clusters can be observed in large  
optical and X--ray samples \citep{plionis02}.  
 
\item {\bf Wavelet coefficients}. 
The mexican-hat wavelet (MHW) is the second derivative of a Gaussian, and it has been  
proposed as an ideal  
filter for compact source subtraction \citep[e.g.][]{cayon00,vielva01}.  
Although the use of the MHW and the number of coefficients (scales) is somewhat arbitrary, 
 we will  
include them just to show that the inclusion of more observables does not pose problems for 
this method.  We take the 3 MHW coefficients at the center of the  
cluster, which produces coefficients that are highly correlated with the second derivative.   
By changing the scale of the MHW we are sampling the cluster at different radii. The three scales  
considered in this work are $s = 0.25, 0.75, 1.66$ arcmin.     
\end{itemize} 

\subsection{Principal Component Analysis}\label{section_PCA} 
 
Principal component analysis (PCA hereafter) has been widely used in the last years as a  
powerful classificator of data sets \citep[e.g.][]{deeming64,teuber79,whitney83,ronen99}.  
For our particular case, PCA has several desirable advantages which can be briefly  
summarized as follows:  (i)  PCA produces an optimal linear combination of the observables that  
maximizes the variance of the linear combination (or {\it projection}), (ii) there is no limit in  
the number of observables, (iii)  it is a {\it non-parametric method}, which means no assumptions  
about the cluster scaling relations are required, (iv) the principal components returned by PCA  
are by definition independent, which simplifies their use in the final computation of the redshift  
of the cluster. \\ 
 
In this paper we will give only a brief description of PCA. The reader is referred to the abundant  
literature about PCA for a more detailed description of the method.\\ 
Let us consider a multivariate data set formed by $N$ observations, each  
observation producing $m$ {\it observables}. In our particular case the observations  
will be  the resolved SZE images  and the {\it observables} will be the  
morphological quantities derived from each one of the images.  
This data set can be considered as an array of $N$ elements in an space of $m$ dimensions. 
That is, each observation has $m$ coordinates.  
Hereafter, we will refer to our data set as the matrix $X_{Nm}$.   
The idea of PCA is that, in many cases, it is possible to reduce the dimensionality of the problem  
without losing any significant amount of information. PCA is specially powerful in those cases where  
there are correlations between some of the observables. In this case, the dimensionality of the problem  
is reduced by {\it projecting} the entire data set over a new orthogonal coordinate system  
which is aligned with the direction of the main correlations in the $\Re ^m$ space.  
The direction of the correlations in  $\Re ^m$ can  
be found by minimizing the sum of distances between the data points and the direction of the correlation.  
However, this is equivalent to maximizing the variance of the data points when projected  
onto the direction of the correlation.  This is what PCA does.

Finding the principle components reduces to an eigenvalue problem (eigenvectors and eigenvalues of the  
covariance matrix $S = X_{Nm} \times X_{Nm}^{\dagger}$). 
The information carried out by each one of the principal components (eigenvectors) is proportional to the  
value of its associated eigenvalue. So, the eigenvector with the highest eigenvalue contains the  
highest amount of information. On the contrary, the eigenvectors associated with the lowest eigenvalues  
does not retain much useful information and they can be dominated by the {\it noise} in the data.  
One can, therefore, consider only the eigenvectors associated with the highest eigenvalues to  
{\it compress} the data set.  In our case, the $m$ observables for  
each cluster will be compressed into $p$ principal components. The criterion to  
chose the value of $p$ is given by the percentage of the total variance retained by the $p$ highest  
eigenvalues.  Usually a good criterion is to retain only those eigenvalues for which the previous  
percentage is  about $90-95 \%$.  For this particular application, we will see in the next section how we  
can retain approximately $\approx 90 \%$ of the variance with only  
the first three principle components.  As we will see, our list of observables is highly redundant!  
 
\subsection{Redshift estimation} 
\label{sect_z_estimation} 
The final component is the probability distribution for the cluster redshift $z$ given the data $d$.  
We use Bayes theorem for this purpose; 
\begin{equation} 
P(z/d) \propto P(z)P(d/z) 
\label{eq_Bayes} 
\end{equation} 
where $P(z)$ is known as the {\it prior}; it  provides the probability of any cluster to be at  
redshift $z$.  
This prior is cosmologically dependent, and it is in principle well defined if one knows the selection  
function of the survey.  For simplicity, we will consider a constant prior in this work; however,  
when estimating cluster redshifts in a real survey where a good estimate of the cluster redshift  
distribution is known (i.e. $P(z)\propto dN/dz$), this information should also be included when  
making cluster redshift estimates. 
The second term, $P(d/z)$, is known as the {\it likelihood} of the data.  In our case, the data are the  
three principal components. Because these components are orthogonal by construction, we model the  
likelihood as; 
\begin{equation} 
P(d/z) = P(pc_1/z)P(pc_2/z)P(pc_3/z) 
\label{eq_z_from_PC} 
\end{equation} 
where $pc_i$ is the $i^{th}$ principal component.   
Each one of this individual probabilities, $P(pc_i/z)$, gives the probability of the observed  
principal component, $pc_i$, to be associated with the redshift $z$.  
 
Computing accurate $P(pc_i/z)$ is absolutely critical, because errors would likely lead to biases  
in the redshift estimates.  The safest approach is a process that we call 
{\it self-calibration}, which requires an observed training set of clusters with  
independently known redshifts.  Such a training set could be arranged by simply carrying out a  
portion of the SZE survey in regions of the sky that have been spectroscopically observed as part  
of the SDSS or 2dF surveys. At high $z$, our calibration method will be limited by the availability 
of identified clusters at those redshifts and a follow up of these clusters will be needed. 
With measured redshifts for some of the clusters, we can compute $P(pc_i/z)$ for each  
principal components over a range of redshifts.  
Here we will model the pdf as a Gaussian with two free parameters, the mean value  
of the principal component at redshift $z$, $\overline{pc_i}(z)$, and its dispersion at the same  
redshift, $\sigma_{pc_i}(z)$. With this form the likelihood for $pc_i$ is 
\begin{equation} 
P(pc_i/z) = e^{-\frac{(pc_i - \overline{pc_i}(z))^2}{2 \sigma_{pc_i}(z)^2}} 
\label{eq_likelihood} 
\end{equation} 
Other probability distributions (i.e. Poisson, $\chi ^2$) could be used,  and if the training  
set were large enough, one could use the histogram (pdf) of the principal components directly.

As a first application of our method, we have applied PCA to 
the toy model of subsection \ref{section_SZE_Observables} with 5 observables; 
total flux, total size, central amplitude, first derivative and second derivative. 
The result is that only the first two resulting principal components are relevant. 
The first one retains 72.46 \% of the total variance and the second one 
retains 27.53 \%. This is not surprising since there are only two independent 
variables in the toy model (redshift and mass). The first PC is dominated by 
the two derivatives and the total size while the second PC is dominated by the 
central amplitude and the total flux. The recovered redshift is unbiased and the errors  
are small ($1 \sigma$ error less than 10 \%). 
 
\section{Application to SZE Images from Hydrodynamical Simulations} 
\label{sec:application} 
 
In this section we will apply our method to SZE images of simulated galaxy clusters. 
But first we provide a brief description of the simulations.  

\begin{figure} 
   \ifthenelse{\equal{\figtype}{EPS}}{ 
   \begin{center} 
   \epsfxsize=8.5cm  
   \begin{minipage}{\epsfxsize}\epsffile{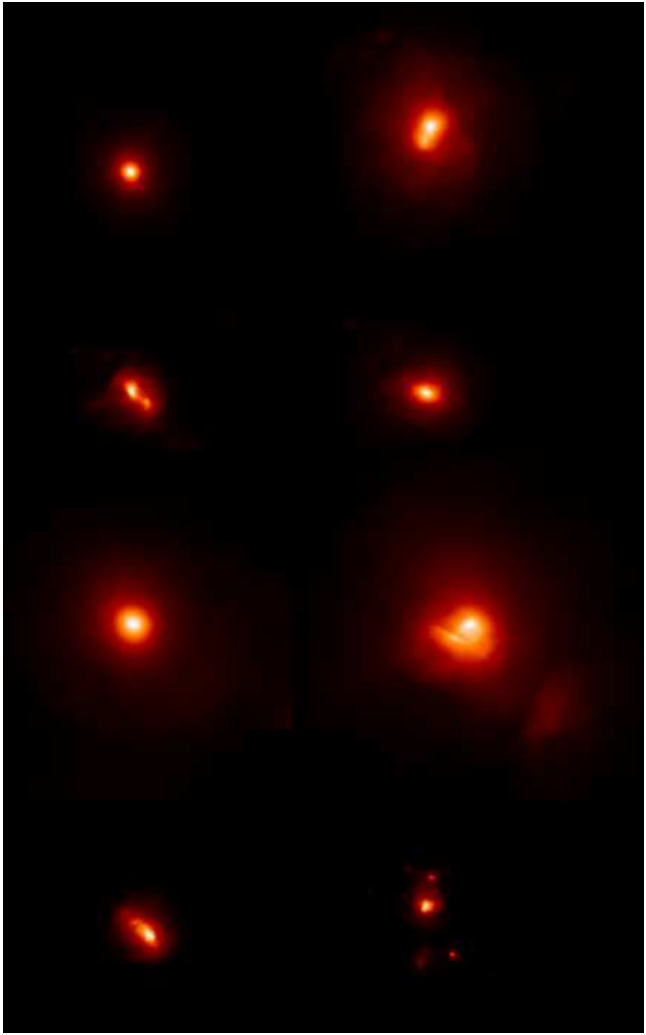}\end{minipage} 
   \end{center}} 
   {\myputfigure{f5.pdf}{3.30}{0.65}{-30}{-30}} 
   \caption{\label{fig_Bunch_Clusters} 
            A sample of simulated clusters with different masses and redshifts.   
            From top to bottom. Upper row, two  
            clusters with the same redshift ($z = 0.25$) but different masses. Next row,   
            two clusters with redshifts 0.75 (left) and 0.5 (right) and different masses.  
            Next row, two clusters  
            with the same redshift ($z = 0.1$) but different mass. Bottom row, two clusters with  
            redshifts 1.0 (left) and 1.5 (right) and different masses.  
            The dimension of this image is 80 arcmin in the horizontal direction and 160  
            arcmin in the vertical direction.   
           } 
\end{figure} 

\subsection{Simulations}\label{simulations} 
 
The clusters were simulated with a combined dark matter and hydrodynamics 
code in a cosmological-constant dominated universe ($\Omega _m = 0.3$, $\Lambda = 0.7$, 
$h=0.7$, $\Omega _b = 0.026$ and $\sigma _8 = 0.928$).  The dark matter was  
modeled with an adaptive particle-mesh method, while the gas was followed  
with an adaptive mesh refinement technique \citep{bryan99,norman99}.   
This grid-based method adds additional meshes in high-density  
regions to obtain high resolution in the central regions of clusters,  
while low-density regions are simulated at low resolution in order to keep  
the CPU requirements modest.  The best resolution so obtained is 16 kpc,  
with generally about 100,000 particles within the virial radius of each  
cluster.  The simulations do not include cooling and star formation due to  
the uncertainty in modelling these processes.  This somewhat reduces the  
realism of the simulations, but because we are only testing the method with  
the simulated clusters, not calibrating it, this should not affect our  
conclusions substantially. 
 
Clusters used in this study are taken from a volume-limited, simulated 
sample and range in mass from 0.7 to $2.0 \times 10^{15} M_\odot$ at 
$z=0$.  Maps of the Compton parameter $y_c$ are generated, some of which 
are shown in Figure \ref{fig_Bunch_Clusters}.  The same clusters are imaged at a variety of 
redshifts, so the various redshift samples are not fully independent (this 
makes our conclusions conservative in the sense that it should decrease 
the observed differences and hence make it more difficult to separate the 
various redshifts than with a real observational sample).  
The clusters are imaged at the following discrete redshifts,  
$z = 0.05, 0.1, 0.25, 0.5, 0.75, 1.0,$ and 1.5.

\begin{figure} 
   \ifthenelse{\equal{\figtype}{EPS}}{ 
   \begin{center} 
   \epsfxsize=8.cm  
   \end{center}} 
   {\myputfigure{f6.pdf}{3.4}{0.60}{-80}{-55}} 
   \caption{\label{fig_3PC} 
            The data set $X_{Nm}$ projected onto the first three principal components.  
            Bluer clusters (larger dots) are low redshift ones, red  
            points (smallest size) are  highest redshift, green/yellow 
            clusters (intermediate sizes) lie in between (intermediate $z$).  
       } 
\end{figure} 
 
\subsection{Observations} 
\label{observables} 
 
We filter the images with a Gaussian filter (FWHM 25 arcsec)  
to simulate the effect of the finite instrument resolution.  
This resolution is achievable with some current  
experiments \citep{pointecouteau01}, but achieving this resolution is not straightforward.  
For instance, it requires a single dish antenna of 30~m diameter like IRAM working  
at $\sim$3~mm to achieve this resolution.  Interferometers can  also produce images with  
these resolutions. 
The instrument sensitivity is included by setting a threshold on the filtered  
Compton parameter images. The sensitivity of the experiment to the SZE signal will  
depend basically on the instrumental noise and the confusion noise (mainly due to  
primordial CMB and point sources). The confusion noise can be reduced with multifrequency  
experiments, which allow partial subtraction of the CMB component.  Higher angular resolution  
observations dramaticaly reduce the point source noise contribution (easily carried out  
by existing interferometers with long baselines).

Here we  assume that we can see  
only the cluster emission above a specific threshold that corresponds to   
$y_c^{th} = 8.0 \times 10^{-6}$.   
This threshold corresponds to our isophote in the isophotal size and flux.  
With this threshold, we lose some of the high redshift ($z = 1.5$) clusters, which after filtering  
have a surface brightness below the threshold, but we can still observe most of our simulated  
high$-z$ clusters. 
  
Our data set $X_{Nm}$ is then a matrix with a number of rows, $N$, equal to the number of observed  
clusters in the survey, and a number of columns, $m = 11$, equal to the number of  
observables for each cluster. 
When using PCA, it is convenient to re-scale the observables in order to make them  
of the same order of magnitude. Here we use the $log$ of the observables and then solve for  
the principle components of the covariance matrix, $S = X_{Nm} \times X_{Nm}^{\dagger}$.   
We find that the first three principle components are responsible for $\approx$90\% of the dispersion  
in our data.   
That is, 3 principle components contain almost all the information within our 11 original observables. 
  
\begin{table} \label{components} 
   \begin{center} 
         \begin{tabular}{|c|c|c|c|c|} 
	 \hline 
	 \hline 
	 & PC1 & PC2 & PC3 \\ 
	 \hline 
	 \hline 
         $\lambda$& 5.15 & 3.56 & 1.03  \\ 
	 \hline 
         Percentage & 46.8 & 32.4 &  9.4 \\ 
	 \hline 
	 \hline 
         Observables &  & Eigenvectors    \\ 
	 \hline 
	 \hline 
         Isoph. Flux& -0.97 & -0.21 & -0.04 \\ 
	 \hline 
         Isoph. Size& -0.98 & -0.11 & -0.05 \\ 
	 \hline 
         Central Amp.& -0.38 & -0.88 & 0.01 \\ 
	 \hline 
          $\partial ^2$ & 0.75 & -0.62 & -0.03\\ 
	 \hline 
          $\partial$ & 0.81 & -0.47 & 0.07 \\ 
	 \hline 
         perimeter& -0.97 & -0.11 & -0.16 \\ 
	 \hline 
         ellipticity& 0.18 & -0.2 & -0.66 \\ 
	 \hline 
         $N_{groups}$& 0.21 & -0.1 & -0.74 \\ 
	 \hline 
          MHW1 (0.25) & 0.62 & -0.76 & -0.02 \\ 
	 \hline 
          MHW2 (0.75) & 0.07 & -0.96 & -0.02 \\ 
	 \hline 
          MHW3 (1.66) & -0.65 & -0.71 & -0.01 \\ 
         \hline 
      \end{tabular} 
      \caption{First 3 eigenvectors of the principle component analysis (columns) and  
               associated eigenvalues (first row)  
               of the 11 observables outlined in $\S$\ref{observables}. The numbers  
               in parenthesis are the typical scales of the MHW's in arcmin.   
               The second row gives the associated percentage 
               for each one of the eigenvectors. The principal components are each  
               a linear combination of the observables; the coefficients of the combination  
               are listed.} 
    \end{center} 
\end{table} 
 
Table~\ref{components} contains the first three eigenvectors with their associated eigenvalues ($\lambda$)  
and percentages. 
The form of the eigenvectors clearly shows which of the 11 observables are the most  
relevant. The first principal component (with the highest eigenvalue), PC1, is dominated by the  
isophotal flux, the isophotal angular size, the perimeter and the first derivative.  
The second principal component  is dominated by the central amplitude, the MHW coefficients  
and the second derivative and the third principal component is dominated by the ellipticity  
and the number of subgroups. 
It is not surprising to see that the flux and size are contributing significantly to the  
most relevant principal component (PC1). On the contrary, the ellipticity and number of subgroups  
only contribute significantly to the third principal component.

\begin{figure} 
   \ifthenelse{\equal{\figtype}{EPS}}{ 
   \begin{center} 
   \epsfxsize=8.cm  
   \begin{minipage}{\epsfxsize}\epsffile{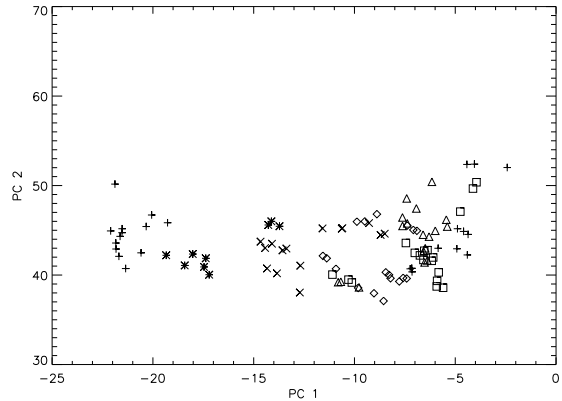}\end{minipage} 
   \end{center}} 
   {\myputfigure{f7.pdf}{3.2}{0.55}{-180}{-30}} 
   \caption{\label{fig_2PC} 
            Projection in the first and second principal components space.  
            $(+)$'s are for $z = 0.05$, $(*)$'s for  $z = 0.1$, (x)'s for $z = 0.25$  
            diamonds for $z = 0.5$, triangles for $z = 0.75$, squares for $z = 1.0$  
            and $(+)$'s again for $z = 1.5$. 
           } 
\end{figure}

In Figure~\ref{fig_3PC} the data set is projected in the space of  
the three principal components. We have used a {\it spectral} scale of colors. That is,  
blue points are clusters at low redshift, green and yellow points are intermediate redshift and  
red points are high redshift. As can be seen, different redshifts are {\it grouped} in  
different regions in this 3D space. This will allow us to discriminate between low, intermediate and  
high redshift clusters. 
Figure~\ref{fig_2PC} contains the projection of the original data set  
in the space defined by the first and second principal components only. The different grouping of  
clusters as a function of their redshift can be also appreciated in this space. 
 
To estimate the redshift we use the expression given in Eqn.~\ref{eq_z_from_PC}, which 
requires the quantities $\overline{pc_i}(z)$ and $\sigma_{pc_i}(z)$ which must be estimated from the  
training set.  In our case the simulated clusters lie at discrete redshifts ($ z = 0.05, 0.1, 0.25, 0.5,  
0.75, 1.0,$ and 1.5), and so we compute $\overline{pc_i}(z)$, and $\sigma_{pc_i}(z)$ only at those  
redshifts. A number of clusters $> 10$ per redshift interval is needed in order to properlly  
estimate $\overline{pc_i}(z)$, and $\sigma_{pc_i}(z)$.   
The total number of clusters in our simulations is $\approx 100$ and there are about  
$10-15$ clusters at each one of the discrete redshifts. Due to this low number of clusters, we  
have taken the training set to be coincident with the total sample of clusters. 
Since we have discrete redshifts in our simulation, we have to interpolate  
$\overline{pc_i}(z)$ and $\sigma_{pc_i}(z)$ for arbitrary redshift.   
In a real survey the situation could be much better if redshifts were available for a larger number of  
clusters with a more continuous distribution in $z$. Then, the training set could be larger  
and without any need of interpolation.  

\begin{figure} 
   \ifthenelse{\equal{\figtype}{EPS}}{ 
   \begin{center} 
   \epsfxsize=8.cm  
   \begin{minipage}{\epsfxsize}\epsffile{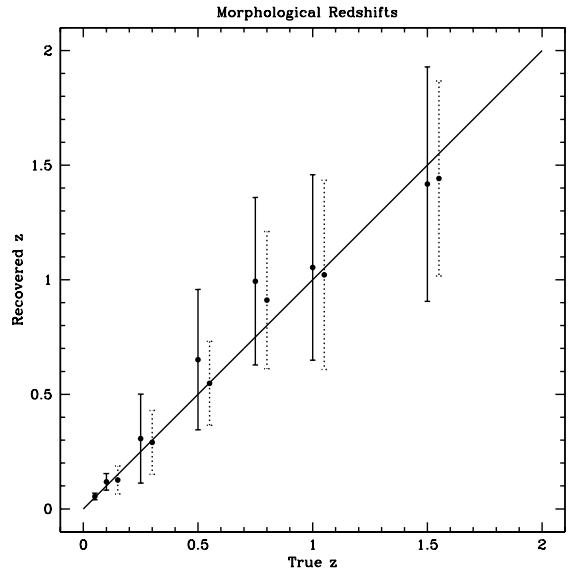}\end{minipage} 
   \end{center}} 
   {\myputfigure{f8.pdf}{3.25}{0.50}{-85}{-30}} 
   \caption{\label{fig_z_true_vs_z_recov} 
            Mean recovered redshift and error bar (dispersion) as a function of redshift.  
            The solid line represents the ideal situation where the recovered redshift  
            equals the true one.  
            For comparisson we also show the correponding recovered redshifts when only  
            the first principal component is used in the Bayesian approach (dotted error  
            bars). The error bars for this case have been displacced 0.05 units in redshift  
            to the right. 
           } 
\end{figure}

\begin{figure} 
   \ifthenelse{\equal{\figtype}{EPS}}{ 
   \begin{center} 
   \epsfxsize=8.cm  
   \begin{minipage}{\epsfxsize}\epsffile{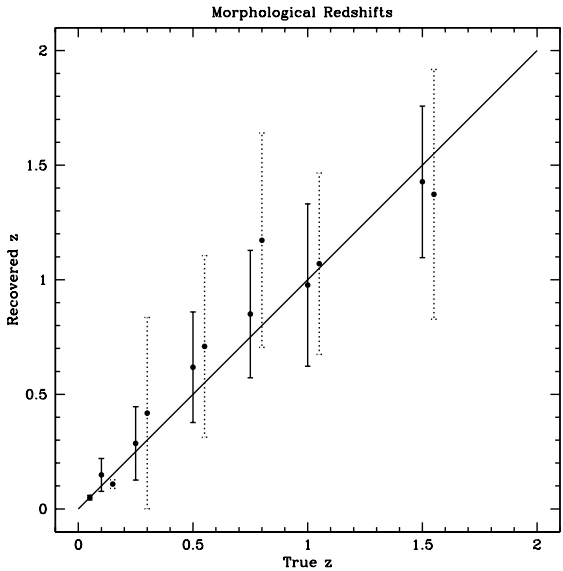}\end{minipage} 
   \end{center}} 
   {\myputfigure{f8.pdf}{3.25}{0.50}{-85}{-30}} 
   \caption{\label{fig_z_true_vs_z_recov_3vs8Obs} 
      Recovered vs true redshift in the case where only three observables: central amplitude, isophotal flux, 
      and isophotal size are considered in the PCA analysis (solid line). 
      This result is comparable with what the one obtained 
      when the 11 observables are considerd but only the first PC was used in the redshift estimation. 
      The dashed lines show the corresponding redshift estimation when the remaining 8 observables are 
      considered in the analysis (first and second derivatives, 3 MHW coefficients, ellipticity, 
      number of subgroups and perimeter) and the central amplitude, isophotal flux, 
      and isophotal size are excluded. The true redshift has been displaced 0.05 to the right to 
      avoid overlapping. Note the good constraints on $z$ obtained by these eight observables at low $z$.
           } 
\end{figure}

Once we have $\overline{pc_i}(z)$, and $\sigma_{pc_i}(z)$ for different redshifts we can apply the  
Bayesian estimator (see Eqn.~\ref{eq_z_from_PC}) for each one of the remaining clusters. 
Figure~\ref{fig_z_true_vs_z_recov} shows the final result of our method. The mean of the  
recovered redshifts follows very well the true redshift of the clusters. The error bars are small  
at low redshifts but they grow larger at higher redshift. 
We also show the result obtained when only the first principal component is considered in the  
analysis (point at $z=0.05$ is not shown). In this case, the error bars and bias are smaller at the  
high redshift interval but they are larger at the smaller redshifts. 
If we look at Figure~\ref{fig_2PC} and we project the points into $PC1$ and $PC2$ we see how  
the projection into $PC2$ is more {\it noisy} in the sense that the overlap of the different  
redshift intervals is stronger than in the projection into $PC1$. This overlap affects the individual  
likelihoods (Eq. \ref{eq_likelihood}), which can show a bimodal or trimodal behaviour specially at  
high redshift (i.e. the individual likelihood for $PC2$ (and $PC3$) has local maxima at different  
redshifts). The addition of the second and third principal components in the analysis adds {\it noise}  
to the z-estimation in particular in the high redshift interval. On the contrary, at low redshifts,  
the second and specially the third principal component (see Figure~\ref{fig_3PC}) show a clear dependence  
with the redshift which helps to better estimate z.  
Consequently the redshift estimation becomes more noisy when we include the second and third principal  
components at high redshift but, in the low-z interval, the second and third principal components  
helps to reduce the error bars. \\
To understand this behaviour, it is helpful to study how different observables contribute 
to the redshift estimation. We have split the list of observables into two groups and 
applied PCA to each group. 
In the first group we include three of the most relevant observables, central amplitude, 
isophotal flux, and isophotal size while in the second group we include the remaining 8 
observables, first and second derivatives, 3 MHW coefficients, ellipticity, number of subgroups 
and perimeter. We compare the results in Figure~\ref{fig_z_true_vs_z_recov_3vs8Obs}. 
The first group renders a $z$-estimation similar to the case where only PC1 is used in the analysis 
(see Figure~\ref{fig_z_true_vs_z_recov}). This is not quite surprising since PC1 was dominated by 
the isophotal flux and isophotal size. 
This result shows that even with a small number (3) of observables it is possible to get 
an estimate of the redshift. However, the other observables also contain information about the 
redshift. This is also illustrated in Figure~\ref{fig_z_true_vs_z_recov_3vs8Obs} (dotted line). 
In this case it is important to note how the 8 additional observables help to reduce the scatter 
in the low redshift interval. Thus it can be useful to include more observables in the analysis to 
reduce the uncertainty. However, the additional 8 observables increase the scatter at higher 
redshifts. \\

Our results show that morphological redshifts are not {\it precise} estimators of the  
cluster redshift, but they are useful providing a first guess that could be critical in planning the  
cluster followup observations to determine photometric or spectroscopic redshifts.  Also, we note that  
the redshift distribution expected for cluster surveys does not contain any sharp features in redshift,  
suggesting that even moderately accurate redshifts like those possible with morphological estimators  
may be sufficient for deriving cosmological constraints.  This clearly deserves further attention. 
 
\section{Conclusions}\label{sec:conclusions} 
 
We have developed a means of estimating galaxy cluster redshifts using only observed SZE properties of  
the clusters. Using a toy model we show how morphological quantities associated to clusters  
may contain redshift information. We also show how modelling of the morphological quantities can lead 
to systematic errors in the redshift estimation. We then propose an alternative method which is 
model independent.
Specifically, we have combined several redshift sensitive SZE observables using a  
standard principal component analysis (PCA).  The PCA led to significant compression, showing that most  
of the redshift information contained in the 11 SZE observables can be expressed in three orthogonal  
linear combinations.  The use of the PCA has several advantages. These include (i) no required  
assumptions about cluster scaling relations, (ii) straightforward to use of direct observables  
(like the isophotal quantities), and (iii) orthogonality of the principal components.  
The method must be {\it calibrated}, and we suggest using a cluster training set that has redshift  
estimates from photometric or spectroscopic means. This training set is required to build the  
likelihoods of the principal components as a function of redshift.  
 
In our analysis we include 11 different {\it observables}: 
isophotal flux, isophotal size, central amplitude, second derivative at the center,  
the mean of the first derivative in the region where the second derivative vanishes,  
the ellipticity and perimeter of the isophote, the number of subgroups above the isophote,  
and three Mexican--hat wavelet coefficients evaluated at the cluster center. 
Principle components were determined, and the first three components had $\approx 90$ \% of  
the variance of the data.  
Application of our redshift estimator using these three components indicates that the method can  
distinguish between clusters at low, intermediate and high redshift. 
 
Although the error bar for a specific cluster redshift is fractionally large, our method should be  
useful for future SZE surveys, providing a preselection of low, intermediate and high redshift clusters.  
This preselection can be used to optimize the optical followup.  Because of the smoothly varying nature  
of the cluster redshift distribution expected in future surveys, it may also be possible to obtain  
cosmological constraints directly with these morphological redshifts. 
As shown in \citet{fan01}, the ratio of the number  
of clusters above and below a given redshift can be a useful cosmological discriminator.  
This kind of analysis could be well suited to our morphological redshift estimates.

A requirement for morphological redshifts is resolved, SZE cluster images.  Our estimates were carried  
out assuming an instrument resolution of 25~arcsec.  This resolution requirement makes our method  
inappropriate for application to clusters detected in the {\it Planck} Surveyor mission, but there  
are several planned interferometric and single dish SZE surveys which could take advantage of our method. 
 
Although in this work we have only considered the case of the SZE, our method can be extended  
to X-ray and optical cluster surveys. The main difference would be that the  
flux in the X-ray and optical bands are inversely proportional to the luminosity distance squared  
and the region of the spectrum observed by a particular instrument also varies with redshift. 
The difference between the luminosity distance and the angular diameter distance is a  
factor $(1 + z)^2$. In general, the X-ray and optical flux is much more sensitive to the cluster redshift  
than is the SZE flux. 
Although the redshift of galaxy clusters in X-rays can be obtained, for some clusters,  
directly from their X-ray spectrum (with typical errors of $\Delta z \approx 0.2$), for many clusters  
with a low SNR the redshift can not be obtained from this method.  Large, planned X--ray surveys will have a  
preponderance of low signal to noise detections, making the use of morphological redshifts (alone 
or combined with photometric redshifts) very promising. 
 
\section*{Acknowledgments} 
This research has been supported by a Marie Curie Fellowship  
of the European Community programme {\it Improving the Human Research  
Potential and Socio-Economic knowledge} under  
contract number HPMF-CT-2000-00967.  JJM acknowledges financial support from the NASA Long  
Term Space Astrophysics grant NAG 5-11415. 
 
\def\apj{ApJ} 
\def\aap{A\&A} 
\def\aaps{A\&A Supp.} 
\def\apjl{ApJ Letters} 
\def\mnras{MNRAS}

\bibliographystyle{mn2e} 
\bibliography{cosmology} 
 
\bsp 
\label{lastpage} 
\end{document}